\documentclass{article}
\usepackage{slashed,feynmf,epsfig,amsmath,amssymb,enumitem}
\usepackage[papersize={8.5in,11in}]{geometry}
\begin{document}
\title{Gravitational Entropy and the Second Law of Thermodynamics}
\author{J. W. Moffat\\~\\
Perimeter Institute for Theoretical Physics, Waterloo, Ontario N2L 2Y5, Canada\\
and\\
Department of Physics and Astronomy, University of Waterloo, Waterloo,\\
Ontario N2L 3G1, Canada}
\maketitle




\begin{abstract}
The spontaneous violation of Lorentz and diffeomorphism invariance in a phase near the big bang lowers the entropy, allowing for an arrow of time and the second law of thermodynamics. The spontaneous symmetry breaking leads to $O(3,1)\rightarrow O(3)\times R$, where $O(3)$ is the rotational symmetry of the Friedmann-Lema\^{i}tre-Robertson-Walker spacetime. The Weyl curvature tensor $C_{\mu\nu\rho\sigma}$ vanishes in the FLRW spacetime satisfying the Penrose zero Weyl curvature conjecture. The requirement of a measure of gravitational entropy is discussed. The vacuum expectation value $\langle 0\vert\psi_\mu\vert 0\rangle\neq 0$ for a vector field $\psi_\mu$ acts as an order parameter and at the critical temperature $T_c$ a phase transition occurs breaking the Lorentz symmetry spontaneously. During the ordered $O(3)$ symmetry phase the entropy is vanishingly small and for $T < T_c$ as the universe expands the anti-restored $O(3,1)$ Lorentz symmetry leads to a disordered phase and a large increase in entropy creating the arrow of time.
\end{abstract}

\maketitle

\section{Second Law of Thermodynamics and Gravitational Entropy}

The second law states that the entropy of a closed system should increase with time and this entropy increase should apply to the whole universe. The 
statistical probability for the entropy $S$ is given by
\begin{equation}
S=k\log{\cal V},
\end{equation}
where $k$ is Boltzmann's constant and ${\cal V}$ is the total volume of phase-space. The quantity ${\cal V}$ is also interpreted as the number of microstates in a given macrostate. The second law tells us that we expect the initial state of the universe at or near the big bang has a vanishingly small entropy, and as the universe expands without any future constraint the entropy monotonically increases. A puzzle that we confront with the second law is why the entropy gets smaller and smaller as we go back further into the past towards the initial big bang. An explanation begs for the existence of a significant physical constraint on the initial state of the universe at or near the big bang.

The state of the universe at the beginning was nearly or completely thermalized, which is observed to be the case for the success of big-bang nucleosynthesis (the predictions of helium and lithium abundance are in close agreement with observation) and the inhomogeneity at the surface of last scattering is of order one part in $10^{-5}$. It can be argued that the entropy of the universe is due mainly to the geometrical structure of spacetime. This structure can be described by the tidal distortion that measures the curvature as well as the distortion due to the presence of matter. 

The tidal distortion can be determined by the Weyl conformal tensor $C_{\mu\nu\rho\sigma}$ and the Ricci tensor $R_{\mu\nu}$, according to the decomposition of the Riemann tensor:
\begin{equation}
R_{\mu\nu\rho\sigma}=C_{\mu\nu\rho\sigma}+K_{\mu\nu\rho\sigma},
\end{equation}
where
\begin{equation}
K_{\mu\nu\rho\sigma}=2R_{\mu\rho}g_{\nu\sigma} -\frac{1}{3}Rg_{\mu\rho}g_{\nu\sigma},
\end{equation}
and where $R={R_\mu}^\mu$. Einstein's field equation:
\begin{equation}
R_{\mu\nu}-\frac{1}{2}g_{\mu\nu}R=\frac{\kappa}{2}T_{\mu\nu},
\end{equation}
introduces the energy-momentum tensor of matter, where $\kappa=16\pi G/c_0^4$ and $c_0$ is the measured speed of light today. The Weyl curvature scalar is given by
\begin{equation}
C^{\mu\nu\rho\sigma}C_{\mu\nu\rho\sigma}=R^{\mu\nu\rho\sigma}R_{\mu\nu\rho\sigma}-2R^{\mu\nu}R_{\mu\nu}+\frac{1}{3}R^2.
\end{equation}

The entropy of matter in the universe is due to the gravitational clumping of matter. As the matter clumps the entropy increases and the maximal entropy occurs for the extreme case of a black hole.  Penrose~\cite{Penrose1,Penrose2,Penrose3} attributed the gravitational entropy of spacetime to the conformal Weyl curvature.

Penrose proposed the Weyl curvature conjecture (WCC) which hypothesizes that the Weyl curvature tensor $C_{\mu\nu\rho\sigma}$ and the Weyl curvature scalar $C_{\mu\nu\rho\sigma}C^{\mu\nu\rho\sigma}$ vanish at or near the big bang. This means that somehow the universe began as a homogeneous and isotropic Friedmann-Lema\^{i}tre-Robertson-Walker (FLRW) spacetime. Whether or not the initial big bang event was described by a singularity (the singularity could be erased by quantum gravity), this strong constraint can be made consistent with the second law.  

All inflation models~\cite{Guth,Linde,Steinhardt} have to explain how enough e-folds of inflation can be expected from a generic initial state of the universe. Aside from the controversial problem of attaching a measure and a probability to the start of inflation~\cite{Carroll}, most if not all inflation models initially assume a sufficiently large patch of spacetime is described by a FLRW geometry. Without the assumption of an initial FLRW geometry an alternative generic state of the universe at the big bang implies that the universe began with a high entropy contradicting the second law of thermodynamics. We are therefore confronted with the question: what is the physical constraint that determines that the universe started with a vanishingly small Weyl curvature and entropy?

\section{Definition of Gravitational Entropy}

Several definitions of gravitational entropy constructed from the Weyl curvature have been proposed, including the simple choice $S=C^{\mu\nu\rho\sigma}C_{\mu\nu\rho\sigma}$. This choice has been criticized~\cite{Wainwright,Rothman} and another choice $S=C^{\mu\nu\rho\sigma}C_{\mu\nu\rho\sigma}/R^{\mu\nu}R_{\mu\nu}$ has been suggested, although this has also been shown to be open to criticism for radiating sources~\cite{Bonnor}. A satisfactory definition of gravitational entropy should satisfy the requirements, (1) It should be non-negative, (2) It vanishes only for vanishing Weyl curvature, $C_{\mu\nu\rho\sigma}=0$, (3) It provides a suitable measure of the anisotropy of spacetime, (4) It yields the Bekenstein-Hawking entropy of a black hole, (5) It increases monotonically describing the growth of structure in the universe. A recent proposal~\cite{Ellis} for the definition of gravitational entropy uses the Bel-Robinson tensor~\cite{Bel1,Bel2}:
\begin{equation}
T_{BR\mu\nu\rho\sigma}=\frac{1}{4}\biggl(C_{\alpha\mu\nu\beta}{{C^\alpha}_{\rho\sigma}}^\beta+{C^*}_{\alpha\mu\nu\beta}{{C^{*\alpha}}_{\rho\sigma}}^\beta\biggr),
\end{equation}
where ${C^*}_{\mu\nu\rho\sigma}=\frac{1}{2}\epsilon_{\mu\nu\alpha\beta}{C^{\alpha\beta}}_{\rho\sigma}$ is the dual of the Weyl tensor and $\epsilon_{\mu\nu\rho\sigma}$ is the fully antisymmetric Levi-Civita symbol,
$\epsilon_{\mu\nu\rho\sigma}= \epsilon_{[\mu\nu\rho\sigma]}$. The Bel-Robinson tensor is symmetric in all indices, trace-free and covariantly conserved in vacuum or in the presence of the cosmological constant $\Lambda$.

A measure of gravitational entropy obtained from the Bel-Robinson tensor has been considered by Pelavas and Coley~\cite{Coley}:
\begin{equation}
S=\int d\tau W,
\end{equation}
where
\begin{equation}
W=T_{BR\mu\nu\rho\sigma}u^\mu u^\nu u^\rho u^\sigma,
\end{equation}
and $u^\mu$ is a timelike unit vector. The entropy measure is non-negative and observer dependent like the density $\rho$ and pressure $p$. It vanishes only when the Weyl curvature tensor vanishes and as demonstrated in ref.~\cite{Ellis}, it satisfies the requirements (1)-(5) listed above.

\section{Spontaneous Violation of Lorentz and Diffeomorphism Invariance}

We propose that the physical mechanism that produces a vanishingly small entropy near the big bang is a spontaneous violation of Lorentz and diffeomorphism invariance. This spontaneous symmetry violation is an essential ingredient in an alternative to inflation theory~\cite{Moffat1,Moffat2}. At the time $t\sim 10^{-23}\,{\rm sec}$ after the big bang the spontaneous violation of Lorentz invariance is accompanied by a large increase of  the speed of light $c\gg c_0$.  At $t\sim 10^{-35}\,{\rm sec}$ the speed of gravitational waves $c_g\gg c_{g0}$, where $c_{g0}$ is the speed today of gravitational waves. The breaking of Lorentz invariance corresponds to the breaking of the homogeneous Lorentz group $SO(3,1)\rightarrow O(3)\times R$ where $O(3)$ is the rotation group and $R$ is a preferred absolute time.  The group $O(3)\times R$ is the symmetry group of the FLRW spacetime with the metric:
\begin{equation}
\label{FLRW}
ds^2=c^2dt^2-a^2\biggl[\frac{dr^2}{1-Kr^2}+r^2(d\theta^2+\sin^2\theta d\phi^2)\biggr],
\end{equation}
with $K=0,+1,-1$ for flat, closed and open models. 

The spontaneous symmetry breaking model~\cite{Moffat3,Moffat4,Bluhm} is based on the action $S_\psi$ given by
\begin{equation}
S_\psi=\int d^4x\sqrt{-g}\biggl[-\frac{1}{4}B^{\mu\nu}B_{\mu\nu}-W(\psi_\mu)-\psi_\mu J^\mu\biggr],
\end{equation}
where $\psi_\mu$ is a vector field, $B_{\mu\nu}=\partial_\mu\psi_\nu-\partial_\nu\psi_\mu$, $W(\psi_\mu)$ is a potential and $J^\mu$ is a matter current density. We obtain the field equation:
\begin{equation}
\label{psiDiffequation}
\nabla_\mu (B^{\mu\nu})-\frac{\partial W(\psi_\mu)}{\partial\psi_\nu}=J^\nu.
\end{equation}

Let us choose $W(\psi_\mu)$ to be of the form of a ``Mexican hat'' potential:
\begin{equation}
W(\psi_\mu)=-\frac{1}{2}\mu^2\psi_\mu\psi^\mu+\frac{1}{4}\lambda(\psi_\mu\psi^\mu)^2,
\end{equation}
where $\lambda > 0$ and $\mu^2 > 0$.  If $W$ has a minimum at
\begin{equation}
v_\mu\equiv\psi_\mu=\langle 0\vert\psi_\mu\vert 0\rangle,
\end{equation}
then the spontaneously broken solution is given by
\begin{equation}
v^2\equiv\psi^\mu\psi_\mu=\frac{\mu^2}{\lambda}.
\end{equation}
We choose the ground state to be described by the timelike vector:
\begin{equation}
\psi^{(0)}_\mu=\delta_{\mu 0}v=\delta_{\mu 0}\biggl(\frac{\mu^2}{\lambda}\biggr)^{1/2}.
\end{equation}
The three rotation generators $J_i\,(i=1,2,3)$ leave the vacuum invariant, $J_iv_i=0$, 
while the Lorentz boost generators $K_i$ break the vacuum symmetry $K_iv_i\neq 0$. The spontaneous breaking of the Lorentz and diffeomorphism symmetries produces massless 
Nambu-Goldstone modes and massive particle modes~\cite{Moffat3,Moffat4,Bluhm}. A preferred proper comoving time $t$ is chosen in the Lorentz symmetry violating phase corresponding to the comoving time in a FLRW spacetime with the symmetry $O(3)\times R$. 

The Weyl curvature $C_{\mu\nu\rho\sigma}$ vanishes in the FLRW spacetime and guarantees that the entropy $S$ vanishes and corresponds to the ordered state of the maximally symmetric conformal FLRW geometry. We can picture the spontaneous violation of the Lorentz symmetry as a phase transition at a critical temperature $T_c$. As the universe expands with $T < T_c$, the transition from the symmetric FLRW state to the less symmetric state correpsonding to the homogeoeus Lorentz group $SO(3,1)$ is accompanied by a large increase in disorder and an accompanying increase in entropy $S$. We assume that no further spontaneous symmetry breaking of spacetime symmetry occurs in the future, so we have an asymmetric time evolution explaining the arrow of time and the second law of thermodynamics.

\section{Conclusions}

The total entropy of the universe in the present day background radiation is for $10^8$ per baryon, in natural units, $10^{88}$, and including the contribution from black holes is of order $10^{101}$ corresponding in natural units to a total phase space volume $\propto 10^{10^{101}}$~\cite{Penrose3}. This huge entropy must be compared to the small or zero entropy that occurs near or at the big bang. A physical mechanism must have occurred in the initial state of the universe to explain this enormous discrepancy in entropy. We identify the mechanism with a spontaneous violation of Lorentz and diffeomorphism invariance, breaking the homogeneous Lorentz group down to the symmetry of the FLRW spacetime with a preferred proper comoving time $t$. This is accompanied by a large increase in the speed of light and gravitational waves solving the horizon and flatness initial value problems in cosmology, and leading to almost scale invariant power spectra and spectral indices (tilts)  $n_s$ and $n_t$ for scalar perturbative density and tensor primordial tensor gravitational waves, respectively.  

\section*{Acknowledgments}

I thank Martin Green and Viktor Toth for helpful discussions. This research was generously supported by the John Templeton Foundation. Research at the Perimeter Institute for Theoretical Physics is supported by the Government of Canada through industry Canada and by the Province of Ontario through the Ministry of Research and Innovation (MRI).


\begin{thebibliography}{10}

\bibitem{Penrose1} R. Penrose, {\it General Relativity: An Einstein Centenary Survey}, S. W. Hawking and W. Israel, eds. Cambridge University Press, 1979.

\bibitem{Penrose2} R. Penrose, {The Emperor's New Mind}, Oxford University Press, 1989.

\bibitem{Penrose3} R. Penrose, Annal. New York Acad. Sci., {\bf 571}, 249 (1989).

\bibitem{Guth} A. H. Guth, Phys. Rev., {\bf D23}, 347 (1981).

\bibitem{Linde} A. D. Linde Phys. Lett., {\bf B108}, (1982).

\bibitem{Steinhardt} A. Albrecht and P. J. Steinhardt, Phys. Rev. Lett., {\bf 48}, 1220 (1982).

\bibitem{Carroll} S. M. Carroll, arXiv:1406.3057 [gr-qc].

\bibitem{Moffat1} J. W. Moffat, arXiv:1404.5567 [astro-ph].

\bibitem{Moffat2} J. W. Moffat, arXiv:1406.2609 [astro-ph].

\bibitem{Wainwright} S. W. Goode, A. A. Coley and J. Wainwright, Class. Quant. Grav. {\bf 9}, 445 (1992).

\bibitem{Rothman} T. Rothman and P. Anninos, Phys. Rev. {\bf D55}, 1948 (1997).

\bibitem{Bonnor} W. B. Bonnor, Phys. Lett., {\bf A122}, 305 (1987).

\bibitem{Ellis} T. Clifton, G. F. R. Ellis and R. Tavakol, Class. Quant. Grav. {\bf 30}, 125009 (2013), arXiv:1303.5612 [gr-qc].

\bibitem{Bel1} L. Bel, C. R. Acad. Sci. {\bf 247}, 1094, (1958).

\bibitem{Bel2} L. Bel, Cahier Phys., {\bf 138}, 59 (1962).

\bibitem{Coley} N. Pelavas and A. A. Coley, Int. J. Theor. Phys.,{\bf 45}, 1258 (2006).

\bibitem{Moffat3} J. W. Moffat, Int. J. Mod. Phys., {\bf D2}, 351 (1993), arXiv:921120 [gr-qc].

\bibitem{Moffat4} J. W. Moffat, Found. Phys., {\bf 23}, 411 (1993), arXiv:9209001 [gr-qc].

\bibitem{Bluhm} R. Bluhm, N. L. Gagne, R. Potting and A. Vrublevskis, Phys. Rev. {\bf D77}, 125007 (2008), arXiv:0802.4071 [gr-qc].

 




\end{thebibliography}
\end{document}